\PassOptionsToPackage{table}{xcolor}
\documentclass[manuscript, nonacm]{acmart}

\usepackage{censor}
\usepackage{multicol}
\usepackage[defaultlines=3,all]{nowidow}

\newcommand{\changed}[2]{{#2}}
\newcommand{\rchanged}[1]{{#1}}

\usepackage{framed}
\newlength{\leftbarwidth}
\newlength{\leftbarsep}
\renewenvironment{leftbar}{\begin{samepage}
    \vspace{0mm}
    \MakeFramed {\advance \hsize -\width \FrameRestore }}{\endMakeFramed
    \end{samepage}
}

\usepackage{amsmath}
\usepackage{wrapfig}
\usepackage{lipsum}
\PassOptionsToPackage{hyphens}{url}
\PassOptionsToPackage{table}{xcolor}

\setcopyright{acmcopyright}
\copyrightyear{2023}
\acmYear{2023}
\acmDOI{XXXXXXX.XXXXXXX}

\acmConference[SIGCHI '00]{ACM SIGCHI 1900}{January 01--05,
  1900}{Earth}

\acmSubmissionID{1662}

\begin{document}

\title[The Effect of Smoothing on the Interpretation of Time Series Data: A COVID-19 Case Study]{The Effect of Smoothing on the Interpretation of Time Series Data:\\
A COVID-19 Case Study}

\author{Oded Stein}
\orcid{0000-0001-9741-3175}
\affiliation{
    \institution{University of Southern California}
    \country{USA}
}
\affiliation{
    \institution{Massachusetts Institute of Technology}
    \country{USA}
}

\author{Alec Jacobson}
\orcid{0000-0003-4603-7143}
\affiliation{
    \institution{University of Toronto}
    \country{Canada}
}

\author{Fanny Chevalier}
\orcid{0000-0002-5585-7971}
\affiliation{
    \institution{University of Toronto}
    \country{Canada}
}

\renewcommand{\shortauthors}{Stein et al.}

\begin{abstract}
\rchanged{
We conduct a controlled crowd-sourced experiment of COVID-19 case data visualization to study if and how different plotting methods, time windows, and the nature of the data influence people's interpretation of real-world COVID-19 data and people's prediction of how the data will evolve in the future.
We find that a 7-day backward average smoothed line successfully reduces the distraction of periodic data patterns compared to just unsmoothed bar data.
Additionally, we find that the presence of a smoothed line helps readers form a consensus on how the data will evolve in the future.
We also find that the fixed 7-day smoothing window size leads to different amounts of perceived recurring patterns in the data depending on the time period plotted -- this suggests that varying the smoothing window size together with the plot window size might be a promising strategy to influence the perception of spurious patterns in the plot.
}
\end{abstract}



  \begin{teaserfigure}
  \centering
  \includegraphics[width=\textwidth]{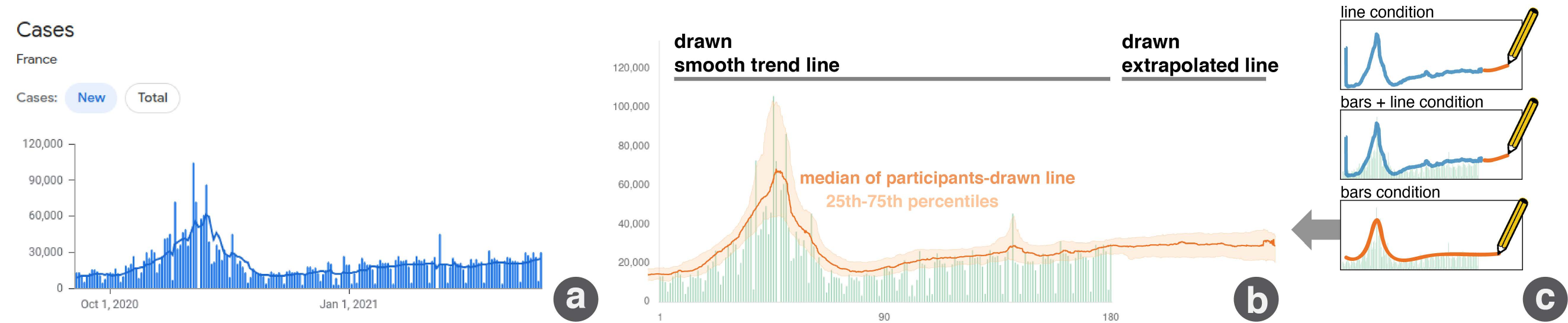}
  \caption{\rchanged{
     COVID-19 case count data is often presented as bars accompanied by a smoothed 7-day average line (e.g. Google News \cite{google-plot-fr} (a)).
In one task of our study, participants were presented with the bars only, and asked to draw a line representing the smooth trend as well as its extrapolation.
    (b) visualizes the participants' drawings for this task on one of our plots.
We also test other visualization conditions (c), where only the smoothed line is presented \emph{(top)}, and where both bars and line are visualized \emph{(middle)}, for which participants were asked to continue the line as they believed the data would continue in the future.
}
  }
  \label{fig:teaser}
\end{teaserfigure} \maketitle

\section{Introduction}

Throughout the COVID-19 pandemic, case count and mortality data visualizations became commonplace and featured daily in news reports and public policy discussions.
\rchanged{
The stakes
are high:
viewers may adapt their behavior (e.g., whether to travel, wear a facemask, get a vaccine) depending on how they interpret the data and how they extrapolate it into the future
\cite{Padilla2022}.
}
\rchanged{
It is well-understood that the choice of visualization method can
dramatically affect people's conclusions \cite{pandey2014persuasive}.
We speculate that
the smoothing techniques employed to remove data-collection artifacts or remove high-frequency fluctuations may simultaneously (and unintentionally) introduce other effects on viewer interpretation.
}

\begin{figure*}
    \centering
    \includegraphics[width=\linewidth]{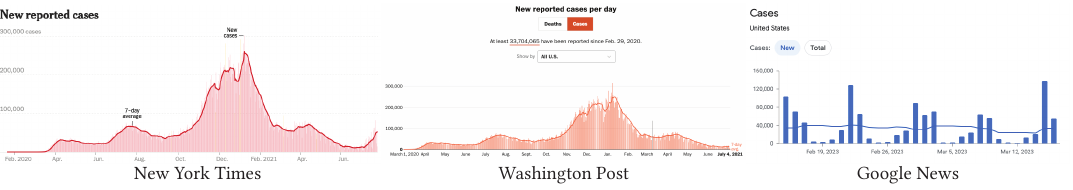}
    \caption{Examples of COVID-19 case visualizations from the New York Times \cite{nytimes-plot}, the Washington Post \cite{washingtonpost-plot}, and Google News \cite{google-plot}.
    In each example, day-by-day case numbers are visualized as bars, and a smoothed case count computed via backward 7-day averaging is plotted as a line.
In this article we research readers' interpretation of this visualization style.}
    \label{fig:newspaperclippings}
\end{figure*}

\changed{whycovid}{
  In this paper, we consider the presentation of daily COVID-19 case count data specifically, as it presents us with a unique opportunity to study human interaction with \rchanged{real-world} time series data.
  For the first time, \emph{billions} of people have regularly looked at and tried to interpret time series (case count data), and used these plots to make important life choices.}
First, we observe and document how
journalism outlets present daily COVID-19 case count data (see ~\autoref{fig:newspaperclippings}).
While some display daily counts directly as bars, many include or instead use a line chart to suggest continuity and employ backward 7-day average to eliminate artificial data-collection patterns (specifically, to account for limited collection during weekends).
As a form of smoothing, averaging also softens anomalies such as large spikes due to one-off dumps of missing data by spreading their influence over multiple days.
Smoothing arises naturally between the tension of desires to be faithful to the collected data, and wanting to remove distracting outliers/details/patterns in order to help readers understand the data.

\begin{figure*}
    \centering
    \includegraphics[width=\linewidth]{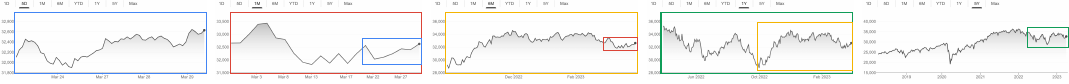}
    \caption{
    Unlike the common presentation of COVID-19 data, stock data is often smoothed
    \rchanged{proportionally}
to the total time duration being visualized \cite{google-stocks}.
``Zooming out'' on the time duration of Google's visualization of
    INDEXDJX
leading up to
    \rchanged{Mar 28, 2023
    highlights this:}
from left (5 days) to right (5 years).
    \emph{Colored boxes added to indicate time relationships. Plots presented in grayscale for clarity.}
    }
    \label{fig:google-stocks}
\end{figure*}

The design space of how to conduct data smoothing for visualization is vast.
Smoothing can be modeled as a low-pass filter in Fourier space, a heat diffusion process, data subsampling, poly-line simplification, or a simple discrete averaging procedure.
For example, common stock price visualizations apply more smoothing the larger the duration of data being displayed (see ~\autoref{fig:google-stocks}).
Stock prices are collected at a frequency much higher than most visualizations' pixel-accuracy, so some smoothing is all but assumed.
In contrast, 
COVID-19 data is mostly collected in daily totals and (especially in the first years of the pandemic) graphical visualizations had pixel-widths much larger than the total number of days.
Thus, visualizing the data losslessly (e.g., as a bar chart) is feasible, and any smooth visualization is a deliberate
choice. 
We observe that
\changed{sevendaybackwardcommon}{
among
news outlets, 7-day backward averaging is a common smoothing choice \emph{regardless} of the number of days displayed
(see~\autoref{fig:newspaperclippings}).
} 
Relative to a fixed pixel-width plot, this
\rchanged{smooths}
more when displaying a smaller total number of days and less for a larger number of days.
Effectively, as the pandemic dragged on the smoothing decreased.

\rchanged{
It is unclear if and how the display of a smoothed line influences people's interpretation of COVID-19 data. To find out, we design and conduct a crowd-sourced study ($n=999$) where
we repeatedly show participants bar charts, line plots or combinations of these, and ask them to interpret the data via answering questions and drawing their own trend lines.
}
Our collected data provides concordant evidence for three primary hypotheses: people were more likely
(i) to identify a recurring patterns when bars are present,
(ii) to identify
\rchanged{an up/down trend}
(as opposed to a constant, horizontal trend) when the 7-day backward smoothed line is present, and
\changed{clarifiedthirdconclusion}
{
(iii) to form a consensus on how the data will evolve in the future when presented with a smooth line in the plot.
}
We also find supporting evidence for two auxiliary hypotheses, and report on main insights from qualitative visual inspection of
\rchanged{the results}.
We conclude with a discussion,
\rchanged{implications for design of smoothing trends in time series} visualization \rchanged{beyond COVID-19 data,} and future \rchanged{work}.

 \section{Related work}

\subsection{COVID-19 Data}
\rchanged{
We conduct a study
on how different visual representations affect the interpretation of COVID-19 case data.
}
Many visualization techniques have been proposed;
the COVID-19 Online VIsualization Collection (COVIC) ~\cite{kahn2022covic} counts around 16,000 unique visualizations.
The potentially harmful role of visualization
\rchanged{is being discussed by experts in personal blogs}~\cite{makulec-blog, correllblog}.
Recent work has discussed the presentation of COVID-19 data from different angles, including design practices
\rchanged{for}
dashboards~\cite{zhang2022visualization}
\rchanged{and}
visual analytics tools~\cite{Bowe2020,leite2020covis,afzal2020visual},
the effectiveness of existing tools ~\cite{Comba2020,Radinsky2022}, the issue of trust in data forecast~\cite{padilla2022multiple}, misinformation on social media~\cite{Lee2021}, and the study of casual visualization used by non-experts~\cite{Trajkova2020}.
To the best of our knowledge, no study has specifically deconstructed one of the most ubiquitous, and seemingly simple representation of daily case counts:
\rchanged{the 7-day} smoothed line accompanied \rchanged{by} raw \rchanged{data}. 

\subsection{Visual Representation of Time 
 Series Data}
 \rchanged{
Visualizing time series data is an essential tool to reason about temporal data; e.g. to allow for visual identification of periodicity in the data, trends, and extraordinary events, and to support the prediction of how data may evolve in the future based on past trends (see survey~\cite{Aigner2023}). 
}
A widespread approach is to display time on the horizontal axis and scalar values as bars~\cite{playfair1822letter,Talbot2014,Skau2015}, lines \cite{playfair2005playfair,Javed2010}, or areas \cite{playfair2005playfair, heer2009sizing} on the vertical axis.

Researchers have studied perceptual properties of graphical techniques used to represent time series data
\rchanged{with a focus}
on comparing performance when conducting visual estimation tasks:
retrieving values from positions and lengths in plots, of different visual variables~\cite{Cleveland1984, heer2010crowdsourcing}, space-efficiency~\cite{heer2009sizing, perin2013interactive, Javed2010} and interaction techniques~\cite{walker2015timenotes,zhao2011exploratory,kincaid2010signallens}.
\rchanged{
In order to make informed decisions, laypeople conceivably draw their own conclusions based on visualizations found in the news (\autoref{fig:newspaperclippings}). In particular, we are interested in the effects of presenting a 7-day smoothed line accompanied or not with raw numbers on people's interpretation of COVID-19 case counts.
Unlike other studies, where performance can be quantified compared to a ground truth, interpreting past COVID-19 data and extrapolating its future do not have well defined ground truth solutions.
}
Instead we ask participants to make a decision which will reveal if and how the visualization affects their understanding of the data shown.
Most relevant to our work is the study of Correll and Heer~\cite{Correll2017}, which explores how people
\rchanged{estimate correlations by performing regression by eye}
in bivariate visualizations. Our work complements and extends this study, by investigating the effect of showing the 7-day averaging line alone, or together with the raw time series data, on people's perception of patterns and prediction of future. 

We are also inspired by prior work which explores drawing as a graphical technique for eliciting prior knowledge about data~\cite{kim2017explaining}; and the works on deceptive visualizations~\cite{pandey2015deceptive,Fan2022}. Like these, our work is motivated by critically evaluating mainstream methods which are used in the news outlets, and the possible impact these may have on people's reasoning and subsequent decision making. 
\changed{algorithmic}{
  As such, we note that while there are standard time series analysis algorithms that can quantify trends, we specifically want to study how the layman interprets plots without any additional annotations, as this is how many COVID-19 plots were presented during the pandemic.
}

\subsection{Smoothing}
\rchanged{
Smoothing is a ubiquitous technique.
It is used to patch holes in image processing~\cite{Perez2003},
fair surfaces in geometry processing~\cite{Desbrun1999},
process noisy data in statistics~\cite{Simonoff1996},
and create visualizations with specific properties \cite{Weinkauf2010,Jacobson2012}.
A wide range of physical phenomena (like heat and diffusion) are modeled using smoothing~\cite[Chapter 20]{Riley2006}.
}

Many smoothing methods are based on the solution of a discretized partial differential equation, such as the diffusion equation. \rchanged{One such example}, the 7-day backward averaging method employed by many outlets to visualize COVID-19 case and death data \cite{nytimes-plot,washingtonpost-plot,sz-plot,google-plot,cnn-plots,economist-conversation,gov-uk-data,npr-plot}, averages the data of the previous 7 days for each day.
This averaging is a finite difference discretization of the diffusion equation
\cite{Thomas1995} with a fixed timestep
\rchanged{independent of} the width of the plot.
There are alternate smoothing methods:
Some publications use a centered 7-day average instead~\cite{sydney-morning-herald-plot}, while others use a moving 7-day or 5-day average \cite{le-monde, timesofindia-plot}. 
For the visualization of financial data, variable smoothing windows are sometimes used (see \autoref{fig:google-stocks}).
\rchanged{Numerical analysis methods for solving the diffusion equation include finite differences~\cite{Thomas1995}, finite elements~\cite{Braess2007}, and finite volume~\cite{LeVeque1992}.}

\rchanged{
There is no uniquely best smoothing method.
}
For some data, like COVID-19 case data, a 7-day smoothing window
\rchanged{can be}
a domain-specific choice to account for the effect of weekends on data collection.
Moreover, additional manual intervention is often employed to address specific outliers \cite{nytimes-anomalies}.
Other editorial decisions might be suitable for different application domains. Because the 7-day backward method has been a widespread method employed by journalists during the pandemic, we are interested in the effects of presenting the resulting smoothed line alone or along with raw counts, compared to raw counts only, on people's interpretation of the data.
 
\section{Research Questions}
\label{sec:researchquestions}

\changed{movedhypearlier1}{
    The visualizations -- viewed by billions -- used for COVID-19 case count data have not undergone systematic scrutiny. Specifically, we do not know how different approaches affect interpretation of the data.
We aim to retrospectively investigate whether the visualization methods utilized by far-reaching news outlets during this unprecedented large-scale delivery influence people's interpretation of the data.
Our main emphasis is on how presenting COVID-19 case counts in their raw form (as bars) and/or in a smoothed form (as a line) affects the perception of past and future trends.
}

\rchanged{
Our \textbf{primary research questions} are:
}
\begin{itemize}
    \item [\textbf{\small RQ1:}] Does the visualization method influence people’s interpretation of the underlying data?
    \item [\textbf{\small RQ2:}] Does the visualization method affect people’s ability to extrapolate data?\footnote{
    \changed{norightorwronganswer}{
      There is no right or wrong answer to the extrapolation task - we want to study \emph{how} the visualization method affects people's extrapolations, not whether they extrapolate correctly.
    }
    }
\end{itemize}

\rchanged{
Our \textbf{auxiliary research questions are:}
}
\begin{itemize}
    \item [\textbf{\small RQ3:}] Does the time window influence people to reproduce 7-day backward averaging when presented with raw case data as bars?
    \item [\textbf{\small RQ4:}] Does the size of the time window influence people's interpretation of the data across visualization conditions?
    \item [\textbf{\small RQ5:}] Does mirroring the data influence people's interpretation of the data across visualization conditions?
\end{itemize}

\changed{whycovid2}{
\subsection*{Why COVID-19 Data?}
  While we are interested in the broader question of the effect of smoothing on the interpretation of general time series data, our work focuses on COVID-19 data.
  This narrow scope has several advantages: (1) familiarity with COVID-19
data is ubiquitous, ensuring that participants understand the data underlying our study's plots;
  (2) COVID-19 data is a rich source of real-world time series data, which means we can perform ecologically-valid studies without relying on synthetic data where the data generation process itself might obscure the result;
  (3) COVID-19 data is a single kind of data, which means that our research is not confounded by mixing different types of data and domains.
This choice narrows our direct conclusions to COVID-19 case counts.
  Our findings may not generalize to other time series data.
We reflect on how our study results can inspire future work on general time series data in \S\ref{sec:discussion}.
}

 \begin{figure*}[t!]
 \centering
  \includegraphics[width=\linewidth]{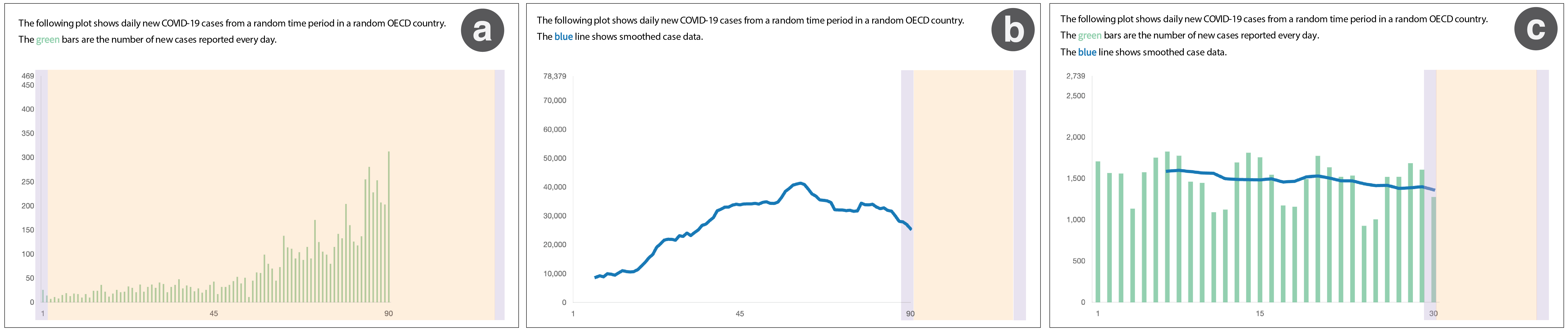}
  \caption{Screenshots of our experiment stimuli, featuring the three \texttt{Vis} presentation methods: (a) \textit{bars}, (b) \textit{line}, and (c) \textit{bars+line}; and different \texttt{TimeWindow} values (i.e. 90 days (a,b), and 30 days (c)). The orange zone corresponds to the drawing zone, which participants were instructed to draw in, starting and ending in the side-ways purple zones (a) to both convey the trend line for the historical data (task \textbf{T4)}, and extrapolate the data (task \textbf{T3)}, or (b, c) to only extrapolate the data, when a trend line was already visualized (task \textbf{T3)}. 
  }
  \label{fig:conditions}
\end{figure*}

\section{Study setup}
\label{sec:studysetup}
\changed{hypmethodmove}{
  We investigate our research questions
  by showing people plots of COVID-19 case data visualized in different ways.
  We asked them to interpret the data via multiple-choice questions and by drawing directly on the plots.
Our hypotheses were formulated and preregistered in advance of data collection.
  For clarity, we postpone their definition until after the description of the study, in \S\ref{sec:hypotheses}.
}

\subsection{Task Design}

\rchanged{We modeled}
our tasks around reading and interpreting COVID-19 case count data \rchanged{presented} visually (see \autoref{fig:conditions}).
For each plot, we asked participants about:

\begin{itemize}
\item \textbf{T1: Observed repeating patterns}, formulated as: \textit{``Do you see any repeating patterns in the plot?''} with possible answers being `yes', `no', `not sure'. 
\item \textbf{T2: Observed trend}, formulated as: \textit{``For the entire plotted time period, the number of cases is primarily trending...''} with possible answers `up', `down', `neither up nor down (constant)'.
\item \textbf{T3: Predicted future}, formulated as: \textit{``Please continue to draw the graph, as you think the data seems to continue''}, with additional details on how to draw and indication that \textit{``You can redraw it as many times as you want.''} (see \autoref{fig:conditions}-b-c).
\changed{linedrawingclarification}{
Participants had to extrapolate the data on the right of the plot by drawing a line with their mouse (the drawing interface is described in \S\ref{sec:procedure}).
}
\item \textbf{T4: Estimated trend}. For only the plots where the smoothed data was not already visualized (see \autoref{fig:conditions}-a), participants were asked to draw the estimated overall trend directly on the plot (in addition to drawing the extrapolated data). The instructions were to \textit{``Please draw a smooth trend line over the entire plot''}. 
\changed{drawclarification}{
For those plots where \textbf{T4} was asked, it was asked in conjunction with \textbf{T3},
and the part of the drawing that went beyond the plot was used to answer \textbf{T3}.
Participants were not explicitly told to mimic 7-day averages.
}
\end{itemize} \rchanged{
For \textbf{T3}/\textbf{T4}, we also asked participants how confident they were in their own drawing, with possible answers `unconfident', `slightly unconfident', `slightly confident', `confident'.
}

\rchanged{
\textbf{T1}, \textbf{T2} and \textbf{T4} are designed to answer \textbf{RQ1}
},
which focuses on reading and interpreting historical data, whereas \textbf{T3} directly addresses \textbf{RQ2}, which pertains to predicting the near future, based on historical data.
\changed{newreftohyp}{
Our hypotheses for how participants will complete these tasks are listed in \S\ref{sec:hypotheses},
and later discussed in \S\ref{sec:discussion}.
}

For the interpretation of historical data, we
\rchanged{choose}
to use both 3-level multi-choice questions (\rchanged{\textbf{T1} \& \textbf{T2};} in the vein of Correll et al.~\cite{Correll2017}) to allow for statistical analyses of people's answers to high level interpretation questions; as well as a free-form drawing on the plot for participants to define their closest fit (similar to Kim et al.~\cite{kim2017explaining}), which allows a finer qualitative analysis of how people perform regression by eye when smoothed lines are not present.
For data forecasting, we
\rchanged{choose}
free-form drawing data collection only, so as to capture participants' closest extrapolation, while guarding against priming effects which pre-defined extrapolated data trajectories would potentially introduce.

We asked participants to solve these tasks on \rchanged{real-world} COVID-19 data.
\rchanged{
The presented raw data is the case number count from a specific country and specific time window (see \S\ref{sec:factors}), pulled from the COVID-19 Data Repository by the Center for Systems Science and Engineering (CSSE) at Johns Hopkins University \cite{JohnsHopkinsCovid}.
}

\changed{whycoviddirect}{
  We explicitly told participants that they were viewing COVID-19 data, and chose against presenting unlabeled data as done by other studies.
  This was done to increase the ecological validity of our study: perception and interpretation of data in real-world scenarios is seldom (if ever) free from confounding factors stemming from domain knowledge and biases.
  There is a spectrum of personal relevance and emotional attachment between data that is completely disassociated from reality, and pandemic data from your own city at the peak of the pandemic. 
We leverage the fact that people emotionally respond to COVID-19 data and are more familiar reading such plots than they would be for other plots.
On the other hand, we are limited by the current date: conducting our study in 2022, this response may have waned, but alas we can not travel back in time to conduct our survey in 2020.
It is also infeasible to present every person with personalized data from their geolocation --- we show generic COVID-19 data, and count on people's general emotional investment.
  We discuss implications of this approach in \S\ref{sec:confoundingfactors}.
}

\subsection{Experimental Factors and Conditions} 
\label{sec:factors}

We were primarily interested in characterizing whether changing the \rchanged{presentation method} of the time series data impacts people's interpretation of this data.
Our main independent variable is:

\textbf{Visualization method} --- \texttt{Vis}: \{bars, line, bars+line\}. Data was presented using one of three visualizations (see~\autoref{fig:conditions}): (i) bars, where the only raw data is shown as bars, at a resolution of one bar (case count) per day; (ii) line, where the only smooth trend data is shown as a line; or (iii) bars+line, where both the raw data and smooth trend line are presented.
\changed{smoothedbarsunsmoothedlines}{
  We do not include smoothed bar plots and unsmoothed line plots, since we did not find them to be common in real-world reporting on COVID-19 data.
}
To create the smoothed line we used 7-day backward averaging:
\rchanged{
the value \rchanged{for} each day is the average of the last 7 days.
Nothing is plotted for the first 6 days.
This approach is inspired by visualization methods observed in the news (see~\autoref{fig:newspaperclippings}).
}

\rchanged{
We also controlled for the following two factors regarding which and how much historical data is presented:
}

\textbf{Historical data} --- \texttt{TimeWindow}: \{30, 90, 180, 360 days\}. We speculate that the temporal window employed in time-series data representations, i.e., how much data from the past is presented to the viewer, might have an impact in their interpretation of the data. We controlled for this time window by presenting data over 30, 90, 180, and 360 days.

\changed{howselected}{
    For each of the 4 \texttt{TimeWindow}s, we selected 3 different countries to source the data.
    We curated individual datasets by randomly selecting different start dates, and retaining the first candidates so as to obtain one dataset where the data at the end of the window trended down, one where it trended up, and one where it trended in no direction at all.\footnote{
    \rchanged{There is no definite mathematical criterion for identifying trends, thus a heuristic was used; this criterion was important to guard against skewed trials, but loosely defined as strict control is unnecessary for our study. See supplemental material for details.}
    }
}

We also wished to explore whether trend observation and data extrapolation would be consistent when the same data was presented with the vertical axis \textit{inverted}.
Mirrored data should theoretically prompt an opposite trend which has the same characteristics (but mirrored) than what the original data would.
We controlled for: 

\textbf{Mirror} --- \texttt{Mirrored}: \{original, mirrored\}. For each of the 12 base datasets, we created its mirrored counterpart by applying a mirroring transform of the $[min_{value},max_{value}]$ interval.
\rchanged{
We included mirrored data to account for \emph{a priori} knowledge bias which can influence perception and interpretation of data~\cite{xiong2019curse}. We suspect that people have expectations as to how the COVID-19 case count data should look like and evolve.
Mirrored data may look more artificial to people. Its inclusion allows us to expose possible confound in our real-world, domain-specific case study.
}

This resulted in a corpus of  
 \texttt{Vis} (3) $\times$ \texttt{TimeWindow} (4) $\times$ \texttt{Mirrored} (2) = 24 unique stimuli used in our experiment.

Our dependent variables include: \texttt{Pattern}, the answer to the multiple-choice question for \textbf{T1}; \texttt{Trend}, the answer to the multiple-choice question for \textbf{T2}; and \texttt{Drawn line}, the precise location of the mouse as well as the drawn pixels for each drawn line, from \textbf{T3} and \textbf{T4}, and \texttt{Confidence}, the recorded self-assessment in how confident participants are in their drawn line as part of \textbf{T3}/\textbf{T4}.

\subsection{Study Platform}
\rchanged{
We recruited participants on Prolific and redirected them to our standalone survey website.
The survey platform was implemented in PHP and HTML/Javascript, results were stored on the institution's MySQL server.
}

\subsection{Procedure}
\label{sec:procedure}

Participants were first prompted with a consent form, and a brief demographics questionnaire to confirm eligibility. Upon completion, each participant was presented with 10 different plots of COVID case data randomly drawn from the corpus of 24 plots.
\rchanged{Participants werw never shown the exact two same plots in succession.}

For each plot, participants were 
first asked to complete
\changed{whatisextreme}{
an engagement question}
(see \S\ref{sec:participants}).
\rchanged{
On the same screen they were prompted to
}
draw a line to complete \textbf{T3} (and \textbf{T4} in the case the plot showed bars only), and answer the multi-choice question asking about their confidence in their drawn line.
Participants had to draw a line in a dedicated zone (orange), with start and end positions restricted to dedicated zones (purple), and answer the question to be able to proceed to the next task.
\rchanged{We} only tracked (and displayed) forward movements during drawing.
\rchanged{On the next screen
participants were asked the two multi-choice questions for \textbf{T1} and \textbf{T2}.}
Once completed, the next trial, with a new plot was presented.
In each multiple-choice question, the order of multiple choice answers was presented in a random order.

\rchanged{
On completion of all trials, participants were invited to share comments in an open text box.
}
\rchanged{
There was no time limit.
}
Participants were \rchanged{paid {\small USD} \$3.60} upon completion for a median completion time of
\rchanged{12m8s}.

\subsection{Recruitement \& Data Collection}
\label{sec:participants}
\rchanged{
The study was approved by the respective Research Ethics Boards / IRBs. We recruited 1,000 participants through the Prolific crowdsourcing platform.
The eligibility criteria included being at least 18 years old, fluent in English, and located in the USA, UK, Ireland, Australia, Canada, or New Zealand.
A participant could only complete the study once.
}

We included an \rchanged{engagement question} to catch answers from inattentive participants and bots:
\textit{``CAPTCHA: What is the largest number listed on the x-axis?''}, i.e., for each plot, participants had to type the length of the plot's time window.
\rchanged{
Trials where participants answered this question incorrectly were excluded (on a by-plot basis, not a by-participant basis).
}
For each plot, we \rchanged{also excluded outliers} based on completion time:
the 2.5\% fastest and 2.5\% slowest trials were excluded from our quantitative analyses.
\changed{norightorwronganswer}{
This was to remove, e.g.\ people who might not have paid enough attention to the task.
}
For the drawing tasks, we specified exclusion of trials where participants indicated they were not confident in their drawing, or which produced some form of invalid drawing due to technical reasons.

Data exclusion criteria, hypotheses, and analysis scripts were preregistered~\cite{osfpreregistration}.
We made some changes to the code to fix mistakes, to add reporting about valid trials, and to account for the flawed initial design of the experiment for secondary hypotheses ($H_E$) -- see final analysis code in the supplemental material.

\section{Hypotheses}
\label{sec:hypotheses}

\changed{hypnew}{
We formulate seven hypotheses to answer the research questions from \S\ref{sec:researchquestions} with our experimental setup.
}

\subsection{Interpreting historical data (RQ1)}
\label{sec:interpreting-historical-data}

\begin{leftbar}
\noindent
\rchanged{
\textbf{$H_A$: When bars are present, people are more likely to spot a recurring pattern.}
}
\end{leftbar}
\rchanged{
We hypothesize this because the bar visualization
allows viewers to visually retrieve the complete original
data, whereas the smooth trend line removes details.
COVID-19 case counts, in particular, are subject to artifacts caused by data collection methods since cases are usually not reported on Sundays, causing a data gap followed by a peak on Mondays. A 7-day backward smoothing trend line presented alone would obfuscate these local artifacts, which the presence of bars would reveal, making viewers more likely to identify a recurring pattern when bars are present.
}

\begin{leftbar} 
\noindent
\rchanged{
\textbf{$H_B$: When the 7-day backward average line is present, people are more likely to spot a non-constant trend.}
}
\end{leftbar}
\noindent
\rchanged{
One of the primary roles of data smoothing is to reduce local noise, and make general trends easier to see, i.e., whether the data is trending up, down, or neither.
Prior work suggests that people tend to down-weight outliers when performing regression by eye~\cite{Correll2017}.
We thus speculate that people might have a harder time confidently identifying a trend without the line, converging to averaging out high-frequency local variations into a constant trend (neither up nor down).
}

\subsection{Extrapolating data (RQ2)}
\label{sec:extrapolating-data}

\begin{leftbar} 
\noindent
\rchanged{
\textbf{$H_C$: People tend to extrapolate less extremely in terms of the slope of the drawn extrapolation when the 7-day backward average line is present than when it is absent.}
}
\end{leftbar}
\rchanged{
We hypothesize that the presence of the smoothed line leads to less extreme extrapolation.
Our intuition is that people tend to be subject to the good continuity Gestalt principle~\cite{wagemans2012century}.
If a seed line to continue from is provided, we believe that people might tend to draw an extrapolated line without dramatic ups or downs, compared to when no such seed is present, which we speculate will result in more extreme data forecasts.
}

\begin{leftbar} 
\noindent
\rchanged{
\textbf{$H_D$: Extrapolated data as drawn by people tend to deviate less across drawers when the 7-day backward average line is present than when it is absent.}
}
\end{leftbar}
\noindent
\rchanged{
We speculate that Gestalt's principle of good continuity~\cite{wagemans2012century} would lead to participants' greater agreement when it comes to predicting beyond an existing smoothed line compared to when no smoothed line is visualized.
}

\subsection{Secondary research questions}
\label{ref:secondaryquestions}

\rchanged{
Our hypothesis for {\small \textbf{RQ3}} is:
}
\begin{leftbar} 
\noindent
\rchanged{
\textbf{$H_E$: People draw trend lines that are less smooth than the 7-day average for small time windows, and lines that are smoother than the 7-day average for large time windows.}
}
\end{leftbar}
\noindent
\rchanged{
We hypothesize that people draw less smooth lines than the 7-day average when the time window is small, and smoother lines than the 7-day average when the time window is large.
}
\rchanged{
One measure of smoothness in calculus is the Dirichlet energy of a function \(E_D(u) = \int_{\Omega} \left| \nabla u(x) \right|^2 dx\) -- the smaller it is, the smoother the function.
We quantify the smoothness of lines using the Dirichlet energy
(see Appendix A),
and thus hypothesize that
\(E_D(\mathrm{7day}) - E_D(\mathrm{drawer})\) increases as the time window increases.
}

\rchanged{
Our hypotheses for {\small \textbf{RQ4}} are:
}
\begin{leftbar} 
\noindent
\rchanged{
\textbf{$H_F$: People are more likely to spot a non-constant trend (up or down) in smaller time windows.}
}
\end{leftbar}
\noindent
\rchanged{
We expect that people believe the number of cases to go up or down more frequently when historical data is presented for a smaller time window.
When less data is available, each daily value carries more weight when interpreting the data, which can result in focusing on local variations more than when data is presented at a the grander scale.
}

\begin{leftbar} 
\noindent
\rchanged{
\textbf{$H_G$: People are more likely to identify non-constant trends in smaller time windows when the 7-day backward line is present.}
}
\end{leftbar}
\noindent
\rchanged{
This hypothesis elaborates on $H_F$.
We hypothesize that the presence of the smoothed line is the main factor influencing identification of nonconstant trends in smaller time windows.
When there is only little data being smoothed, the smoothed line is more likely to be perceived as going into one direction or another.
}

\rchanged{
In the interest of space, we do not include our hypotheses for {\small \textbf{RQ5}} here, since the results end up being inconclusive.
See supplemental material for details.
}

\section{Results}
\label{sec:results}

\rchanged{
We collected 999 valid Prolific submissions on Nov 2, 2022,
yielding 8,960 usable answers to individual plots (some of the contributors did not complete the survey due to technical difficulties). 
After applying pre-registered data exclusion criteria, valid trials remaining for analysis were $5849$ for RQ1, and $2714$ for RQ2  (most usable answers were discarded because of the engagement questions, and for RQ2, lack of confidence in the drawn line).
See the breakdown of valid trials per condition in Appendix~\ref{app:additional-tables}.
}

We report on our statistical analyses, organized per hypothesis.
We follow an estimation approach, drawing inferences from graphically-reported point and interval estimates~\cite{cumming2005inference, dragicevic2016fair}.
The 95\% confidence intervals (CIs)
are computed using bootstrapping~\cite{diciccio1996bootstrap}. Where relevant, we also report $p-$values or linear regression analysis outcomes. 

\begin{figure}
\begin{minipage}[b]{0.48\linewidth}
\centering
\includegraphics[width=\textwidth]{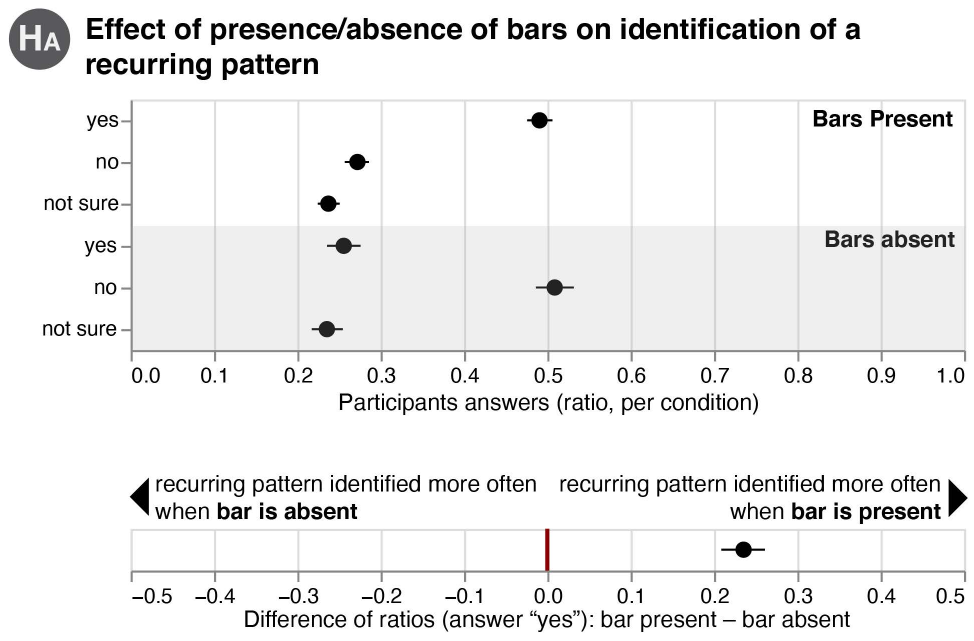}
  \caption{Summary of answers to \textbf{T2}.
  Top: ratio of answers per condition.
  Bottom: Difference of ratios when participants reported seeing a pattern. Error bars are 95\% bootstrapped CIs.}
  \label{fig:hypothesis-A}
\end{minipage}
\hspace{0.2cm}
\begin{minipage}[b]{0.48\linewidth}
\centering
  \includegraphics[width=\textwidth]{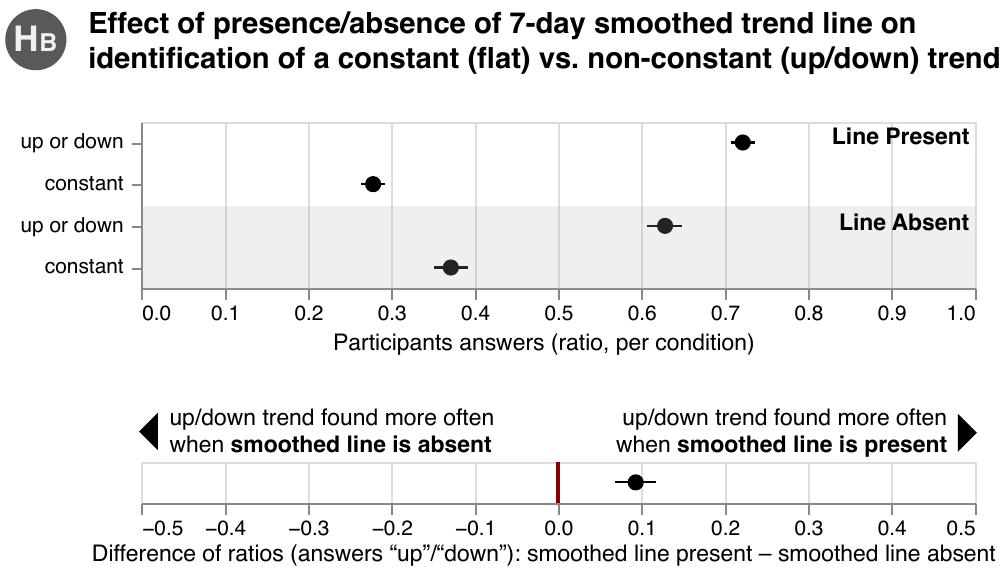}
  \caption{Summary of answers to \textbf{T1}. Top: ratio of answers per condition. Bottom: Difference of ratios when participants reported that the general trend for the historical data was going up, or going down. Error bars are 95\% bootstrapped CIs.}
  \label{fig:hypothesis-B}
\end{minipage}
\end{figure}

\subsection{Hypothesis $H_A$}

\autoref{fig:hypothesis-A} shows a summary of responses to \textbf{T2}:
\rchanged{
Do people spot a recurring pattern?
The top plot shows the proportion of answers relative to the total number of trials where bars representing the raw case counts were displayed (\textit{bars} and \textit{bars + line} together) or not (\textit{line}).
The bottom chart contrasts the difference in proportion of trials where participants identified a recurring pattern between conditions when bars were present and when absent.
}

We find strong evidence in support of the hypothesis:
When bars are displayed, participants were more likely to see a recurring pattern than when
\rchanged{bars were absent}
(overall 23\% more trials {\small $CI:[21,26]$}; $\chi^2 (2, N=5849) = 383.94, p < .001$).

\subsection{Hypothesis $H_B$}
\autoref{fig:hypothesis-B} shows a summary of responses to \textbf{T1}:
\rchanged{
Is there a non-constant trend (up or down) or a constant trend over the entire plotted time?
}
The upper plot shows the percentage of trials for each set of conditions (line present vs. line absent) where participants noted a constant vs. non-constant trend.
The bottom plot \rchanged{contrasts} these conditions.

We find supporting evidence: participants were more likely to identify a trend going up or down when the 7-day backward average line was present than when it was absent. An overall 9\%  {\small $CI:[7,12]$} more trials were identified as non-constant when the smooth line was visualized; $\chi^2 (2, N=5849) = 52.63, p < .001$.

\subsection{Hypothesis $H_C$}
\changed{hc-elaborated}{
To quantify how participants extrapolate data,
}
we fit a straight line to each participant's drawn
\rchanged{line},
and compared the magnitude of the slope in plots where the smoothed line was present to plots where it was absent.
\autoref{fig:hypothesis-C} shows that the difference in slope magnitude is not swaying in either direction for most datasets (i.e. CIs cross the zero line).

Our results are inconclusive. We find an overall difference around zero (0.01, {\small $CI: [-0.00,0.03]$}), between the magnitude of the slope of drawn extrapolated lines when the 7-day smoothing line was visualized and that when
\rchanged{it}
was absent.

\subsection{Hypothesis $H_D$}
\begin{figure}
\begin{minipage}[t]{0.48\linewidth}
\centering
 \includegraphics[width=\textwidth]{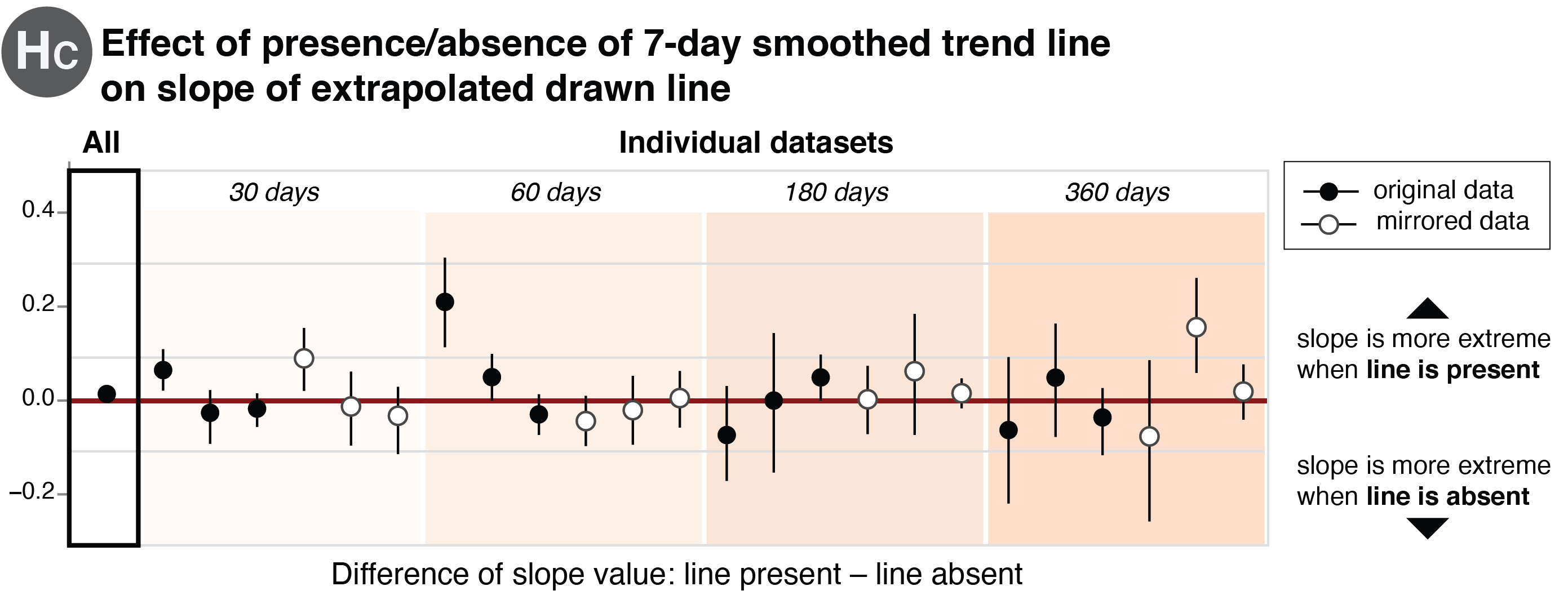}
  \caption{Difference of slope between trials where line is present and absent, for each unique dataset (and mirrored counterparts). Error bars are 95\% bootstrapped CIs.}
  \label{fig:hypothesis-C}
\end{minipage}
\hspace{0.2cm}
\begin{minipage}[t]{0.48\linewidth}
\centering
    \includegraphics[width=\textwidth]{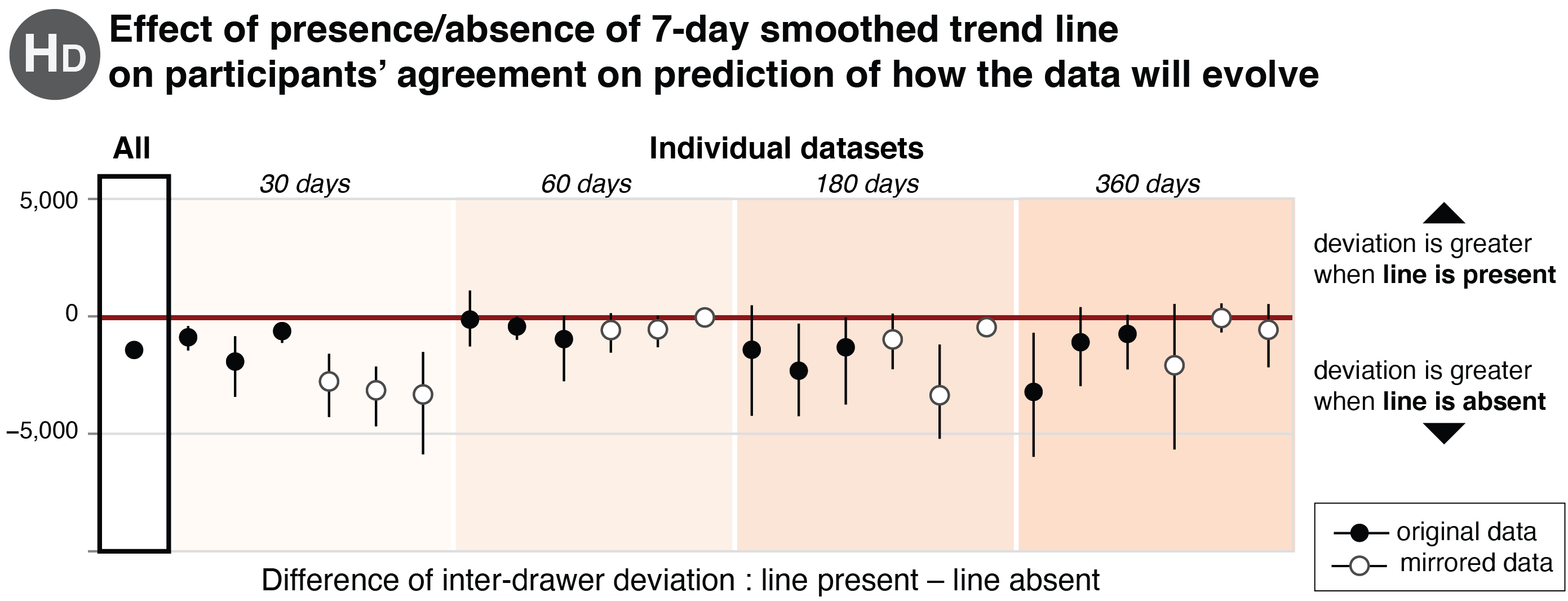}
  \caption{Difference of intra-drawer deviation between trials where line is present and absent, for each unique dataset (and mirrored counterparts). Error bars are 95\% bootstrapped CIs.}
  \label{fig:hypothesis-D}
\end{minipage}
\end{figure}

To investigate this hypothesis, we integrate the standard deviation of all participants' drawn lines for each plotting method and each dataset.
In \autoref{fig:hypothesis-D} we show the difference between these standard deviations for trials where the participants were presented with the 7-day backward average line and trials where the participants were only shown bars.
We find that, overall, the participants drew lines that are more similar to each other when the line is present than when it is absent, although looking at individual datasets, the difference is not significant (CIs cross the zero line).

We find weak evidence for the hypothesis:
The overall deviation is greater
\rchanged{absent the smoothed line}
(the aggregated point estimate and confidence interval are below
\rchanged{zero}:
-1362,  {\small $CI: [-1708,-1051]$}), but we can not completely accept the hypothesis, since the effect is small and for most individual datasets, we do not obtain a significant result.

\subsection{Hypothesis $H_E$}
Our results are inconclusive.
We find some Pearson correlation between time window size and \(E_D\) difference (\(r=0.136\)), but when separated into non-mirrored (\(r=-0.030\)) and mirrored (\(r=0.264\)), this effect seems restricted to the mirrored data.
\rchanged{
We cannot explain this, and suspect that we are capturing a different effect (plots in supplemental material).
}

\subsection{Hypothesis $H_F$}

\begin{wrapfigure}[5]{r}{110pt}
\centering
\vspace{-20pt}
\hspace{-9pt}
\includegraphics[width=0.92\linewidth]{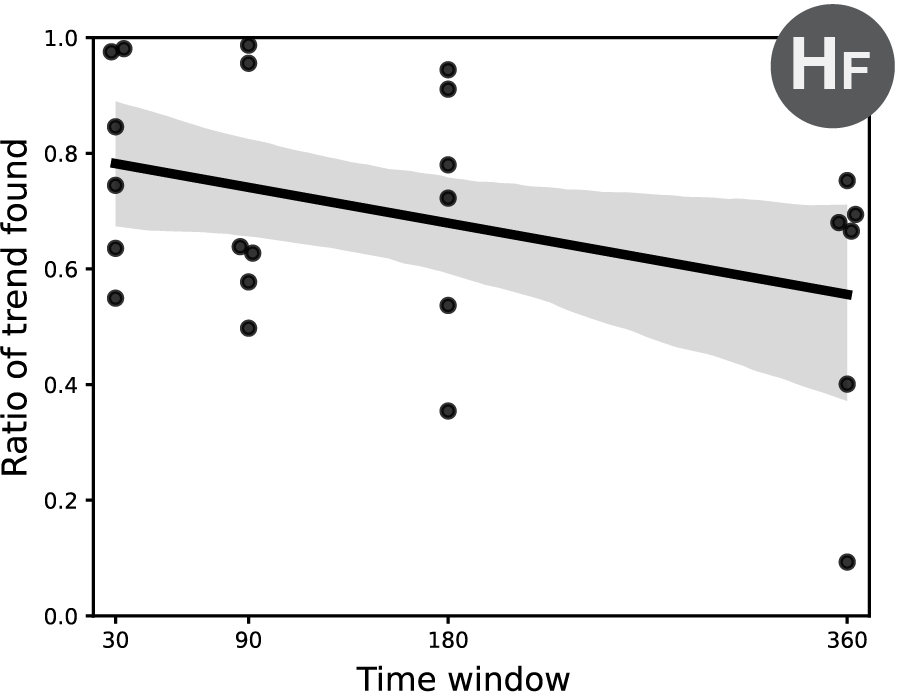}
\end{wrapfigure}
The inset figure shows the aggregated ratio of trials which participants indicated they saw a constant trend, for each set of 6 datasets per time window value.
We find some supporting evidence for this hypothesis: the Pearson correlation ($r = -0.393$) between window length and trend ratio suggests that the larger the time window, the least likely do people interpret the data as going up or down.\hfill \break

\subsection{Hypothesis $H_G$}

\begin{wrapfigure}[5]{r}{130pt}
\centering
\vspace{-20pt}
\hspace{-9pt}
\includegraphics[width=\linewidth]{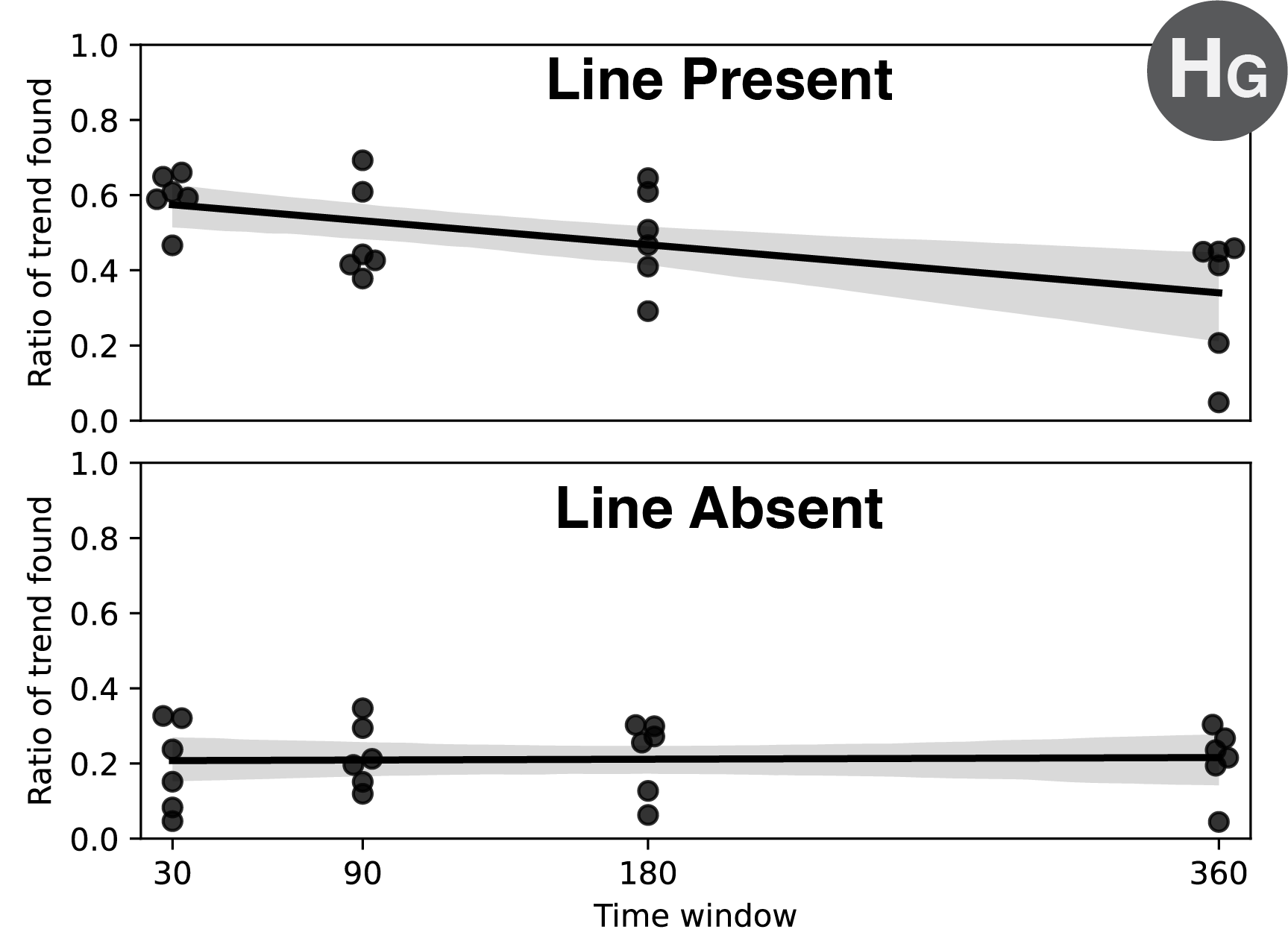}
\end{wrapfigure}
\textit{$H_G$.}
The inset figure compares the overall responses (ratios of trends found to go up or down) aggregated for each unique stimulus for different time windows when the 7-day backward line was present or absent.
We find supporting evidence for this hypothesis. People tend to find non-constant trend in smaller windows when the line is present (Pearson's correlation $r = -0.592$), but the window size does not seem to play a role when the line is absent ($r = 0$).

\section{Qualitative analysis}
\label{sec:qualitative}
In this section, we visually inspect
\rchanged{participant drawings}.
As described in \S\ref{sec:procedure}, each participant drew their own line for each plot of data they were shown (extrapolation only if the smoothed line was present, and over the entirety of the plot and into the future if presented with bars only).
To visualize participants' drawings, we register a vector graphic polyline generated from their recorded mouse movements over a screenshot of the pristine plot
\rchanged{
(accounting for pixel offsets of the drawing area overlay over the displayed plot).
}
Since the plotting tool is displayed slightly differently on each participant's computer due to differences in operating systems and browsers, discrepancies of a few pixels between what the participants saw and the data we recorded could have occurred.

\begin{wrapfigure}[8]{r}{118pt}
\centering
\vspace{-20pt}
\hspace{-9pt}
\includegraphics[width=0.92\linewidth]{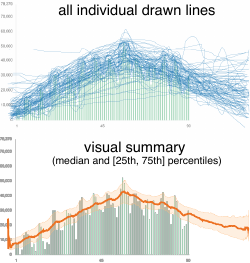}
\end{wrapfigure}
We aggregate all drawn lines \emph{(inset, top)} by drawing percentiles
\emph{(inset, bottom)}.
\changed{visualsummaryexplained}
{
We resample each drawing uniformly along the horizontal axis with 1000 samples, and then draw the median as a thick orange line.
The area between the 25th and 75th percentile is shaded in a light color to highlight the variability of drawings.
This is overlaid
}
over a screenshot of the plot (see also \autoref{fig:teaser}).

We proceed by highlighting qualitative observations about select interesting plots.
The full collection of generated drawings is available in the supplemental material.

\begin{figure}
    \centering
    \includegraphics[width=\linewidth]{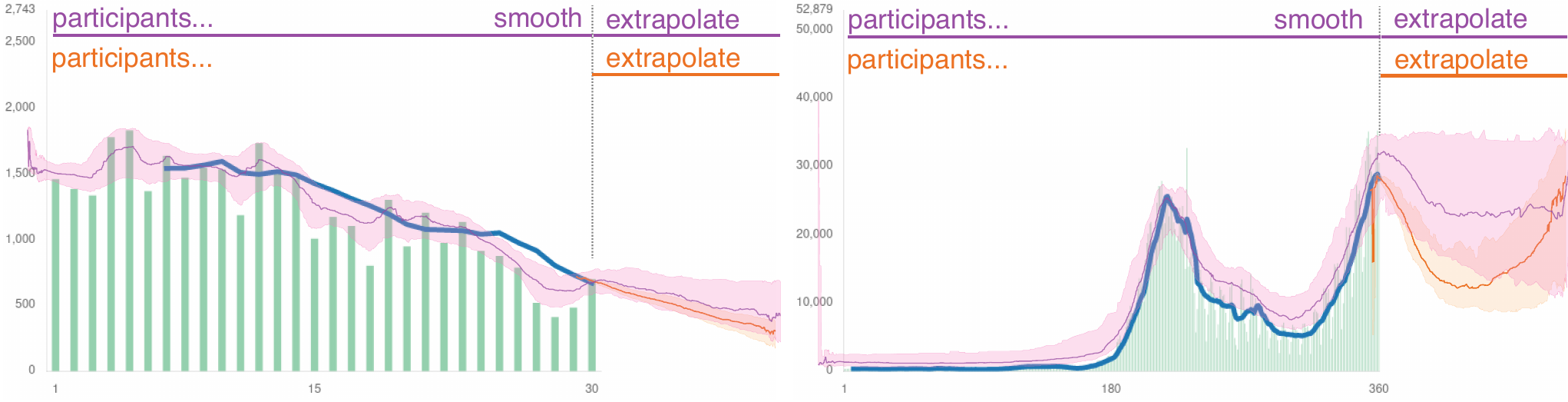}
    \caption{Overlay of two different trials
for a 30-day dataset (left) and a 360-day dataset (right).
    \rchanged{
    In the orange trial, participants are presented with green bars and a blue smoothed line, and asked to extrapolate the line.
    In the purple trial, participants are presented with bars only and asked to smooth \& extrapolate the data.
    }
}
    \label{fig:estonia-poland}
\end{figure}

\subsection{Participant smoothing vs.\ 7-day backward averaging}
\label{sec:participant-vs-seven-day}
By looking at the regions in \autoref{fig:estonia-poland} where drawers had to smooth raw data
(\emph{not} the extrapolation part),
\rchanged{
we can compare the 7-day backward average line (blue) with the drawers' own smoothing work (purple line and region).
}
We observe that in the 30-day plot participants drew a line that is less smooth than the 7-day backward average, and clings closer to the green bars.
In the 360-day plot, participants drew a line that is smoother than the 7-day backward average.
We speculate that this is because drawers chose the smoothness of their line based on how it visually appears to them with respect to the plot window size, and not with respect to the number of days that are displayed.
\rchanged{This}
leads to \emph{undersmoothing} compared to the 7-day backward average line on small time windows, and \emph{oversmoothing} on large time windows.
\rchanged{
Looking at the purple shaded region, we notice that participants were more likely to draw above the top of the green bars than below them.
}
There seems to be an aversion to draw lines
\rchanged{going}
through the green bars.

\subsection{Drawing the entire line vs.\ only extrapolating}
\label{sec:drawing-entire-line-vs-extrapolating}
\autoref{fig:estonia-poland} also shows the differences between the drawers' extrapolation behavior when presented with the raw bar data and the 7-day backward smoothed line vs.\ their behavior when presented only with bars.

\begin{figure}
\begin{minipage}[t]{0.48\linewidth}
\centering
  \includegraphics[width=\textwidth]{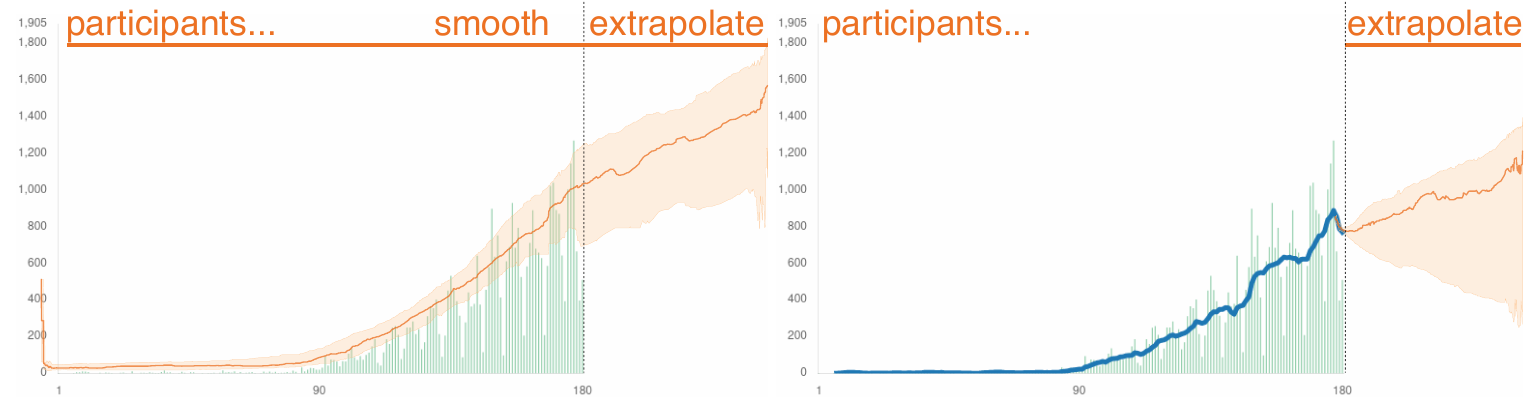}
    \caption{
    \rchanged{
    The same raw data presented in two ways:
    Unprocessed, one green bar per day, drawers are asked to smooth and extrapolate the data (\emph{left});
    and bars as well as a smoothed line, drawers are only asked to extrapolate (\emph{right}).
    }
    }
    \label{fig:extrapolate-vs-all}
\end{minipage}
\hspace{0.2cm}
\begin{minipage}[t]{0.48\linewidth}
\centering
      \includegraphics[width=\textwidth]{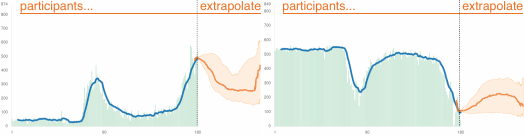}
    \caption{The same real-world data presented to participants as-is \emph{(left)} and mirrored \emph{(right)}.
    \rchanged{
    Drawers are shown raw data as well as a smoothed line, and asked to continue the line on the right.
    }
    }
    \label{fig:mirrored-vs-regular}
\end{minipage}
\end{figure}

\begin{wrapfigure}[5]{r}{90pt}
\centering
\vspace{-18pt}
\hspace{-9pt}
\includegraphics[width=\linewidth]{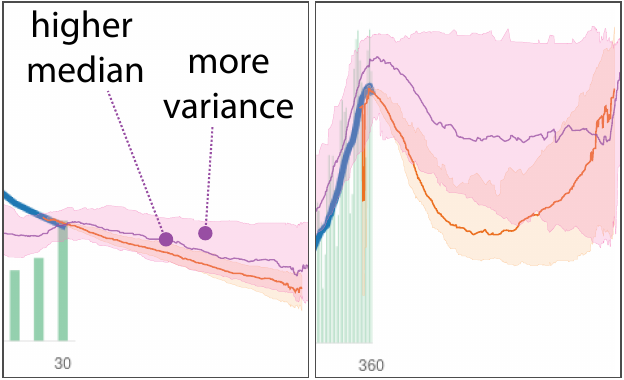}
\end{wrapfigure}
In both examples we can see that drawers believe that future case numbers will be higher when they were not shown the 7-day backward average line \emph{(inset, purple)} compared to when \rchanged{it}
was present \emph{(inset, orange)}.
In the 30-day plot \emph{(inset, left)} this effect is very mild, maybe because of the gentle slope of the raw data at the end of the plot.
In the 360-day plot \emph{(inset, right)} this effect is strong -- the drawers intuit that case numbers will be much higher in the future if not presented with the
\rchanged{smoothed}
line.
Drawers also have a wider spread in drawn lines (the width of the purple vs.\ the orange shaded region) -- the 7-day backward average line leads to more consensus among drawers.

\begin{wrapfigure}[5]{r}{90pt}
\centering
\vspace{-8pt}
\hspace{-9pt}
\includegraphics[width=\linewidth]{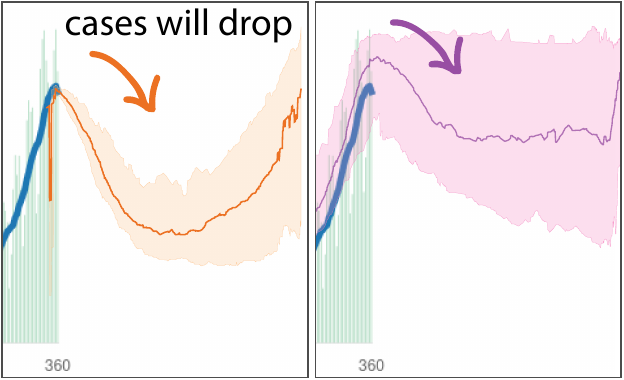}
\end{wrapfigure}

Another interesting effect that can be observed with both sets of drawers in
\rchanged{
this specific}
360-day plot is that they believe the increase at the end of the plot is a temporary peak that will decrease again.
If presented with the 7-day backward average line, however, they believe that after the end of the peak there is yet another peak in the future \emph{(inset, left)} -- drawers are repeating the patterns of previous peaks.
While there might be a semantic reason for drawers to believe that the increase at the end of the plot is part of a peak that will decrease again (due to the nature of new case data, which can not increase forever), they have no knowledge of whether there will be peaks in the future or not -- they are merely repeating the previous pattern of the 7-day average line they were presented with.

\begin{wrapfigure}[4]{r}{90pt}
\centering
\vspace{-8pt}
\hspace{-9pt}
\includegraphics[width=\linewidth]{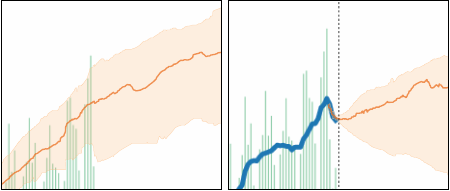}
\end{wrapfigure}
In \autoref{fig:extrapolate-vs-all}, where the raw data is increasing during the entire plot, drawers that were only presented with bars all seem to agree that the number of cases will continue to rise.
The drawers that were provided with the average line as a starting point for extrapolation were split, possibly because of the sharp kink in the line right at the end of the plot \emph{(inset, right)} -- while most drawers also seem to think that the number of cases will continue to increase
(albeit less so than the lineless drawers), many of them believe that the sharp kink is not an anomaly, but actually the top of a peak that signals a decrease in cases after the plot ends.
This effect seems to override the better consensus among people presented with the \rchanged{smoothed} line that we observed in \autoref{fig:estonia-poland}.
\rchanged{
The effect does not occur in all plots (see supplemental material).
}

\begin{figure}
    \centering
    \includegraphics[width=\linewidth]{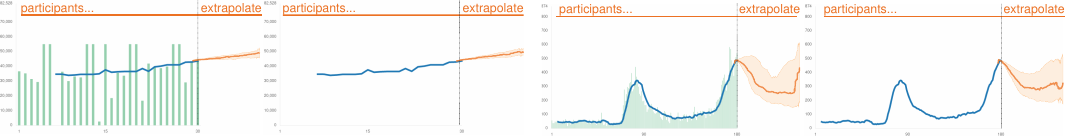}
    \caption{Drawers' extrapolation behavior when presented with a 7-day backward average line and raw data vs. only the 7-day line.
}
    \label{fig:only-lines}
\end{figure}

\subsection{The effect of mirroring}
\begin{wrapfigure}[5]{r}{52pt}
\centering
\vspace{-12pt}
\hspace{-5pt}
\includegraphics[width=\linewidth]{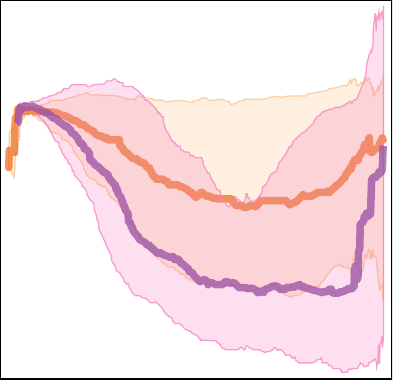}
\end{wrapfigure}
\autoref{fig:mirrored-vs-regular} shows that, when the COVID-19 case data is mirrored, drawers do not always produce a mirrored extrapolation line.
\changed{mirroredvsregularinset}{
The extrapolated data with the mirroring undone (user drawings mirrored back) is overlaid in the \emph{inset} with the data resulting from non-mirrored prompts in purple and mirrored prompts in orange.
}
We speculate that whether the plot ends with a high case number or whether it ends with a low case number, and whether the data at the end of the plot trends upward or downward affects extrapolation behavior beyond mere mirroring -- the semantic difference between an increase and a decrease in cases changes people's interpretation of the plot.
In
\rchanged{this specific case,}
both mirrored and non-mirrored plots the median drawer repeats a pattern from earlier in the plot -- an increase towards the end for the non-mirrored data (as seen in the discussion of \autoref{fig:estonia-poland}), and an increase followed by a decline for the mirrored data.
While 75\% of drawers predict an immediate dip in the non-mirrored data, less than 75\% of drawers predict an immediate increase in the mirrored data (the orange region has a dip on top in the non-mirrored data, but a flat bottom for the mirrored data) -- the pattern effect does not affect plots that end with a high case number and plots that end with a low case number to the same extent.

\subsection{Line \& bars vs.\ only line}
In \autoref{fig:only-lines} we compare showing bars and the 7-day backward average line to showing only the line.
In the first row,
\rchanged{no large}
difference can be observed.
When the blue line is almost constant at the end of the plot, drawers continue it with a small slope and similar amounts of consensus whether bars are present or not.
The second row (same data as in \autoref{fig:mirrored-vs-regular}) shows an example with a more extreme slope at the end.
Here we can see some differences:
\rchanged{The bars lead}
to less extrapolation consensus, an effect also observed in our quantitative analysis of \(H_D\) -- the bars lead some drawers to predict a small increase at the end of the plot.
The difference is not as striking as it is when comparing only bars to bars \& line though (as in \autoref{fig:extrapolate-vs-all}).

 \section{Discussion \& Conclusion}
\label{sec:discussion}

\subsection{What do our results mean?}
\textbf{Plotting the raw data as bars increases the likelihood that people recognize the data to exhibit a recurring pattern.} Far fewer patterns were recognized in the absence of raw data. While we did not explicitly ask participants to identify \emph{which} pattern they saw, the results for {\small \textbf{RQ1}} lead us to think that viewers were able to recognize the spurious 7-day pattern inherent to all raw COVID-19 data.
\changed{falsepatterns}{
We suggest that leaving out raw bar data from the plot successfully hides potentially irrelevant or false patterns from readers, albeit at the cost of data provenance transparency.
}

\textbf{Presenting a 7-day average smooth line helps people see an overall trend in the data.} We found evidence that people were more likely to see an overall trend upward or downward when presented with the average line. We speculate that it is hard for people to decide whether there is an upward or downward trend through regression by eye from raw data alone, and that the smoothed line helps them interpret the plot.
Note that we intentionally refrain from making a determination whether people see actual trends inherent in the data or whether they see spurious trends, since this is difficult to determine with any certainty for COVID-19 data. Our findings just imply that showing a smooth line makes people more prone to see a general trend up/down.

\textbf{Visualizing the average smooth line increases consensus on data forecast.} Our results for {\small \textbf{RQ2}} suggest that including a smoothed line in a plot leads to more consensus among readers on how the data will develop in the future.
We conjecture that adding a smoothing line to a time series plot helps inform participants about future implications of data.
The qualitative analysis in \S\ref{sec:drawing-entire-line-vs-extrapolating} supports this conjecture, but adds a caveat to it:
extremal events present in the smoothed line can lead to significant qualitative differences in reader's extrapolation behavior.

While our analysis of \(H_E\) did not produce results strong enough to confidently state that people undersmooth
\changed{oversmoothundersmooth}{
compared to 7-day backward averaging
}
for small time windows and oversmooth
\changed{oversmoothundersmooth}{
compared to 7-day backward averaging
}
for large windows, our qualitative analysis in \S\ref{sec:participant-vs-seven-day} does suggest that there is something to this effect that should be further investigated.
We found that, for some plots, people follow the noisy raw bars more closely for small time windows (were few bars are prominently visible), but deviate more from them for large time windows (were individual bars are harder to discern).
The 7-day smoothing line does not take the size of the time window into account.

Our results for \(H_F\) and \(H_G\) lead us to surmise that showing a lot of data in a time series plot can hinder the readers' ability to spot an upward or downward trend in the data.
There is possibly a sweet spot where the time window is large enough so that readers do not spot spurious trends, but small enough so they can still identify salient trends -- this is an interesting avenue for future work.
This effect seems to vanish when the smoothing line is absent (which might also point towards the line helping people spot trends in the underlying data).

\subsection{Confounding factors}
\label{sec:confoundingfactors}

We did not consider a number of confounding factors that could have contributed to our results.
Since we showed participants COVID-19 case data, preconceived notions about COVID-19 have the potential to influence people's reaction to the data~\cite{xiong2019curse,kim2017explaining}.
\changed{confoundingcovidreveal}{
  Our original intent was to use the fact that participants would be familiar with COVID-19 plots due to their presence in their daily lives to retrospectively investigate how they interpret plots that were mainstreamed in the news.
  Future work might compare how the revelation of this fact influences the result.
}

We had initially introduced mirrored data to our study to generate more stimuli, and to see whether the direction of a trend matters (i.e., would people give reverse answers to all questions about mirrored data).
Our mirrored experiments were not very conclusive, and we had difficulty interpreting them.
This might be because mirrored data is often readily apparent as such with COVID-19 data.By including these, we gain in generalizability of our findings, but we also observed signals in our analyses suggesting that other biases associated with assumptions about the data domain may be at play, potentially a combination of the negativity and fear instincts~\cite{rosling2016factfulness} and availability heuristic~\cite{curt1984primacy}.

The date of the survey can also influence people's perception of the data.
We conducted the survey in November 2022 -- while this was no longer the peak of the pandemic for the countries we surveyed, the worst parts of the pandemic were still fresh on the minds of the participants.
As time goes on, the reduced prominence of COVID-19 in people's lives can change their interpretations of COVID-19 data.
Lastly, the context in which people read time series plots can affect their interpretation.
Our participants had to complete their survey in front of a computer using a mouse.
People who read plots on the go, on their phone, or while otherwise distracted might interpret them differently.

\changed{confoundingrealdata}{
We did not control for properties of the real-world data we pulled (i.e. steepness, overall variance, etc.)
Properties of the underlying data might have influenced our results.
Since our data was randomly picked from real COVID-19 data, we are confident that it displays similar properties to any COVID-19 plot one might encounter in real life.
}

\changed{linebasedthinking}{
  Because we asked participants to draw lines as one of our tasks, we might have inadvertently encouraged line-based thinking and biased their interpretation of line plots.
  Future work might separate the line-drawing tasks from the other tasks, and present them to different people.
}

\subsection{How to capture people's thoughts: lessons learned}
\label{sec:lessons-learned}

In order to task people to identify trends, one could present them with pre-drawn trend lines \cite{Correll2017}.
Instead, we let users draw their own trend lines, interacting with the plot directly using their mouse.
There are advantages to this:
Participants have a lot of freedom to communicate their thoughts to us precisely.
Since drawn data is \rchanged{rich and continuous, }
we can apply \rchanged{many} mathematical analysis tools to it, and visualize it in informative ways.
For the extrapolation task, presenting participants with pre-drawn choices brings with it a danger of priming, which we can avoid by letting participants extrapolate by drawing.
There are also disadvantages:
Recording drawn data poses more technical challenges than simpler multiple choice.
We had to impose a variety of restrictions on the drawing task to ensure valid input data.
Our tool forced drawers to
\rchanged{start and end their drawings in specific regions.}
We also forced them to only draw forward -- this restriction ended up causing further problems, as the lines of people moving their mouse backwards while drawing were constrained to not go backwards, causing large jumps in y-value with only tiny advances in x-value (likely contributing to our difficulties with \(H_E\)).
Additionally, each participant experiences a slightly different drawing experience due to differences in
\rchanged{software \& hardware},
which can cause noise in the collected data.

\subsection{Limitations of COVID-19 data}
\label{sec:presentingcoviddata}
\changed{nopandemicquestions}{
We did not take full advantage of the fact that we were presenting participants with COVID-19 data.
Our questions sought to investigate visualizations of smoothed time series data in general and were thus worded to be domain-agnostic.
Future work might ask domain-specific questions that are more relevant for the communication of pandemic data.
}

\changed{seeingreference}
{
Looking at labeled graphs on topics for which people have preconceived notions is known to bias people's interpretation of the data \cite{xiong2022seeing}.
This affects our study, since our participants were told that they are looking at COVID-19 data, a topic which has been very present in everyone's life in the year leading up to the study.
While this allows for more ecologically-valid results for our specific case, this choice could limit the generalizability of our conclusions.
}
\changed{coviddatanotpersonalized}{
While our work does use COVID-19 data in order to motivate participants to engage with the plots similarly to how they engaged with similar plots while reading the news, we did not match up participants to data that is personally relevant to them, e.g., data from their location.
Future work might present every participant with plots directly relevant to their city, and study this emotional response more directly compared to the response for a random location.
}

\section{What to consider when smoothing}
\label{sec:what-to-take-into-consideration}
We also intend this article to serve as a starting point for designing visualizations for smoothed time series data. While our work focuses on COVID-19, it has broader implications.
\rchanged{
In this section we discuss what to take into account when smoothing time series data, as well as alternative smoothing methods.
}

\subsection{Visualization of smoothed data}

Whether to display a smoothed line or not, and, if it is displayed, whether to display it in conjunction with raw data, has an effect on people's interpretation of the data.
If the data has known periodic anomalies that readers should ignore (i.e. weekends have an impact on data collection for COVID-19 that is not reflective of the underlying case data), then smoothing is invaluable -- our study provides evidence for the fact that people do not account for these anomalies on their own.
For such cases, a smoothed line engineered specifically to remove anomalies should be considered.
Should one include the raw data when presenting time series?
Our results suggest that the presence of bars makes people see more repeating patterns -- whether or not these patterns should be communicated depends on the circumstances~\cite{Rong2017}.
\changed{trueorfalsepatterns}{
  If a designer does now know whether their data has false patterns that should be removed, or true patterns that should be highlighted, our study does not offer strong guidance.
  We suggest including smoothed visualization if very high-frequency noise is present, and setting the strength of the smoothing relative to the size of the display window.
}

The presence of a smoothed line seemingly helps people form a consensus about the future evolution of time series data.
A smoothed line in a visualization can thus make it easier to communicate future predictions -- one should however be aware of the dangers of pigeonholing readers.
We have also seen evidence suggesting that the presence of a smoothed line makes it easier for people to spot underlying trends in the data, making the communication of trends easier.

Another interesting takeaway from our article is that people seem to have less consensus on how to interpret extreme events
\changed{whatisextreme}{
(i.e., stark outliers like the peak in \autoref{fig:teaser})
}
than on how to interpret other data.
When presenting data with extreme events, one might thus consider explaining such events via additional communication, such as annotations~\cite{law2020characterizing}.
More elaborate, extrema-preserving smoothing methods could be a good solution 
\cite{Weinkauf2010,Jacobson2012}.

\subsection{Alternative Smoothing Methods \& Outlook}

Even though we did not experimentally evaluate alternatives to 7-day backward averaging in this work, we discuss other smoothing methods here.
Their experimental evaluation is future work.

\begin{figure}
    \centering
    \includegraphics[width=\linewidth]{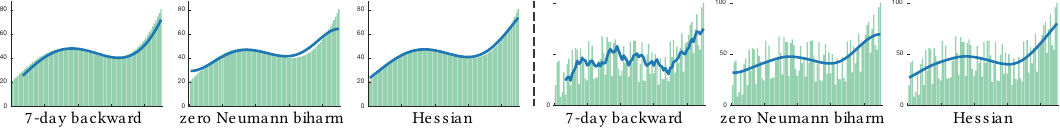}
    \caption{
    Smoothing synthetic data to illustrate the behavior of different methods.
    Left: non-noisy data.
    Right: noisy data.
}
    \label{fig:synthetic-smoothing}
\end{figure}

Seven-day backward averaging can be understood as a particular discretization of the mathematical diffusion equation in one dimension, \(\frac{\partial u}{\partial t} = k \frac{\partial^2 u}{\partial x^2}\).
The diffusion strength \(k\) is fixed in the 7-day averaging method and can not be changed.
Thus, when data is very noisy (or the dataset contains lots of timesteps), the effect of the smoothing diminishes.
Methods that do not suffer from this are variable-timestep finite element methods for the diffusion equation \cite{Braess2007} or the higher-order biharmonic diffusion equation \cite{Jacobson2012,Stein2018}
(see \autoref{fig:synthetic-smoothing}).

The 7-day backward average also suffers from lagging behavior.
Because this method is backward-looking, it tends to artificially move events to the future.
This effect is especially visible when smoothing data that is not noisy (see \autoref{fig:synthetic-smoothing}).
One can choose different discretization methods for the diffusion equation that avoid lagging behavior, such as centered 7-day averages that take into account data from the past \emph{and} the future (some news outlets employ these \cite{sydney-morning-herald-plot}).

When smoothing data one must consider the edges of the plot -- where there is \emph{not} enough past (or future) data to average.
If there is crucial information to be communicated at the edges, truncating the smoothed line is not an option.
In 2D/3D geometry processing, smoothing techniques apply various boundary conditions that determine behavior at the edges.
Picking a boundary condition allows us to plot smoothed data up to the very edge. Popular boundary conditions for the biharmonic diffusion equation include the zero Neumann condition (data should be flat at plot edges) \rchanged{and} the Hessian conditions (data should be linear at plot edges), which both correspond to different scenarios from the physical simulation of elastic systems \cite{Stein2018}.
\autoref{fig:synthetic-smoothing} demonstrates smoothing with these boundary conditions.

We demonstrated that the simple and popular 7-day backward average smoothing method has significant effects on the interpretation of plots.
Given the many available other smoothing techniques, we are optimistic that future studies of these methods could reveal further effects of smoothing on the interpretation of time series data.

\begin{acks}
This work was supported in part by the Swiss National Science Foundation’s Early Postdoc.Mobility fellowship.
This work was supported in part by a grant from NSERC (RGPIN-2018-05072)
This research was funded in part by NSERC Discovery (RGPIN–2022–04680), the Ontario Early Research Award program, the Canada Research Chairs Program, a Sloan Research Fellowship, the DSI Catalyst Grant program and gifts by Adobe Inc.
We thank Yvonne Jansen and Souti Chattopadhyay for valuable comments which helped improve the manuscript.
\end{acks}

\bibliographystyle{ACM-Reference-Format}
\bibliography{perception-of-smoothing}

\appendix
\appendix

\section{Computation of the Dirichlet Energy}
\label{app:dirichlet-computation}

We compute the energy in the evaluation of \(H_E\) as follows:

\begin{enumerate}
\item We normalize the x-axis and y-axis data.
\item We eliminate all intervals between points on the x-axis that are smaller than \(10^{-4}\) to prevent numerical issues in calculations down the line.
\item We eliminate the highest-frequency elements of each line -- this allows us to only focus on non-smoothness deliberately introduced by drawers, ignoring mouse tracking error and involuntary movements.
This frequency elimination happens via the eigendecomposition (using \texttt{scipy}'s \texttt{eigsh}) of the finite element piecewise linear Laplacian operator on the plot domain \cite{Braess2007}.
For each line, we only use the eigenfunctions corresponding to the smallest \(n-7\) eigenvalues to compute the Dirichlet energy, where \(n\) is the number of days in the raw data.
\item We compute the Dirichlet energy of the processed line using the eigenfrequencies.
If \(u\) is the line to process as a function from the x-coordinates to the y-coordinates, and \(\lambda_i, v_i\) are the eigenvalues and eigenfunctions of the Laplacian described above, then the eigenfrequencies are defined as
\begin{equation*}
    u_i = \int_{\Omega} u(x) v_i(x) dx,
\end{equation*}
where \(\Omega\) is the x-axis of the plot.
The Dirichlet energy is then
\begin{equation*}
    E_D(u) = \sum_{i=1}^{n-7} \lambda_i u_i^2
\end{equation*}

\end{enumerate}
The implementation can be found in the supplemental material code.

The preregistered version of \(H_E\)'s analysis featured a different way to compute the Dirichlet energy.
We would subsample the drawn line with a fixed number of samples, and then compute the Dirichlet energy of this subsampled line.
This would lead to undesired behavior due to small mouse movements made by the participant during the drawing task, and due to small imprecisions in the recording of mouse positions.
We believe that a Dirichlet energy computed like this does not capture the intention of our original research question, as we conjecture that such small movements are probably not intentional non-smoothness introduced by the drawer due to their interpretation of the data.
Thus we redesigned this analysis to cap the maximum frequency that goes into the computation of the smoothness measure.
We include the results from the earlier version of the analysis in the supplemental material.

\section{Additional Tables}
\label{app:additional-tables}

In this section, we list additional tables that break down the valid and invalid trials for each condition.
Table~\ref{tab:RQ1-valid-trials} lists the numbers for RQ1, and Table \ref{tab:RQ2-valid-trials} lists the numbers for RQ2.

\begin{table}[h]
\small
{\renewcommand{\arraystretch}{1.2}
\centering
\makebox[\linewidth][c]{
\rowcolors{1}{white}{gray!20}
\begin{tabular}{ccc|r}
 \multicolumn{1}{l}{bars} & \multicolumn{1}{l}{bars+line} & \multicolumn{1}{l|}{line} & \multicolumn{1}{l}{total} \\ \hline
 1972 & 1902 & 1975 & 5849
\end{tabular}
}
}
\caption{Number of valid data points included for analysis for RQ1.}
\label{tab:RQ1-valid-trials}
\end{table}

\begin{table}[h]
\small
{\renewcommand{\arraystretch}{1.2}
\centering
\makebox[\linewidth][c]{
\rowcolors{1}{gray!20}{white}
\begin{tabular}{rr|ccc|ccc}
\rowcolor{white}
 & & \multicolumn{3}{c|}{not mirrored} & \multicolumn{3}{c}{mirrored} \\
 \multicolumn{1}{l}{\#days} & \multicolumn{1}{l|}{ID} & \multicolumn{1}{l}{bars} & \multicolumn{1}{l}{bars+line} & \multicolumn{1}{l|}{line} & \multicolumn{1}{l}{bars} & \multicolumn{1}{l}{bars+line} & \multicolumn{1}{l}{line} \\ \hline
 30 & 1 & 42 & 51 & 60 & 40 & 50 & 49 \\
 30 & 2 & 59 & 65 & 69 & 40 & 36 & 45 \\
 30 & 3 & 37 & 57 & 54 & 27 & 63 & 44 \\
 90 & 1 & 39 & 37 & 43 & 48 & 44 & 44 \\
 90 & 2 & 23 & 26 & 27 & 42 & 28 & 38 \\
 90 & 3 & 40 & 50 & 34 & 30 & 42 & 21 \\
 180 & 1 & 36 & 45 & 28 & 38 & 42 & 43 \\
 180 & 2 & 34 & 26 & 31 & 28 & 31 & 19 \\
 180 & 3 & 37 & 32 & 29 & 48 & 37 & 41\\
 360 & 1 & 33 & 19 & 37 & 28 & 20 & 22 \\
 360 & 2 & 26 & 24 & 20 & 28 & 21 & 24 \\
 360 & 3 & 37 & 50 & 45 & 41 & 32 & 38 \\
\end{tabular}
}
}
\caption{Number of valid data points for each stimulus included for analysis for RQ2. Total is 2714.}
\label{tab:RQ2-valid-trials}
\end{table}

\end{document}